\documentclass[12pt]{article}

\newcommand{\beq}{\begin{equation}}
\newcommand{\eeq}{\end{equation}}
\newcommand{\beqa}{\begin{eqnarray}}
\newcommand{\eeqa}{\end{eqnarray}}
\newcommand{\beqar}{\begin{eqnarray*}}
\newcommand{\eeqar}{\end{eqnarray*}}

\def \la {\langle}
\def \ra {\rangle}
\def \up {\uparrow}
\def \down {\downarrow}

\input epsf

\begin{document}

\title{  Entanglement from the vacuum}

\author{
          Benni Reznik \\
          School of Physics and Astronomy \\
          Tel Aviv University \\
          Tel Aviv 69978, Israel.\\
e-mail:reznik@post.tau.ac.il
}
\maketitle

We explore the entanglement of the vacuum of a relativistic field
by letting a pair of  causally disconnected probes
interact with the field.
We find that, even when the probes are initially non-entangled,
they can wind up to a final entangled state.
This shows that entanglement persists between disconnected
regions in the
vacuum.
However the probe entanglement, unlike correlations,
vanishes once the regions become sufficiently separated.
The relation between entropy, correlations and entanglement is
discussed.


\section{\bf Introduction }

The Hilbert space of two subsystems contains a
subclass of entangled states that manifest unique
quantum mechanical properties.
As was highlighted by Bell \cite{bell},
the correlations between observables measured separately
on each subsystem can be ``stronger''  than
the correlations predicted by any local classical models.

Having the causal structure and locality built in,
relativistic field theory offers a natural framework for
investigating
entanglement.
Here we will consider the
entanglement of the relativistic vacuum state, which as we shall
shortly recall, has a role in both the Hawking black-hole
radiation
\cite{hawking}
and  Unruh acceleration radiation effects \cite{unruh}.
It is known that field observables at space-like separated points
in vacuum are correlated.
For massless fields in 3+1-D these correlations decay
with the distance,  $L$, between two points as $ 1/L^{2}$.
These correlations by themselves however do not imply
the existence of quantum entanglement, because
they can in principle arise as classical correlations.
However, a number of studies provide evidence that
the vacuum is indeed entangled \cite{unruh,werner,clifton}.
In the Rindler quantization,
one spans the Hilbert space of a free field by direct products
of Rindler particle number states  $|n,1\ra$ and $|n,2\ra$
with non-vanishing support confined within
the two complementary space-like separated wedges $x<-|t|$ and
$x>t$,
respectively.
It then turns out \cite{unruh}
that the Minkowski vacuum state
can be expressed as an entangled Einstein-Podolsky-Rosen (EPR)
like state
 $\sim \sum_n \alpha^n
|n,1\ra|n,2\ra$ for each mode.
However does entanglement persist when the regions
are separated by a finite distance? We will examine this
question,
but also emphasize that in this case of separated regions the
relation
between entanglement and entropy breaks down.

In a somewhat different framework, of algebraic quantum
field theory, it has been argued  \cite{clifton}
that indeed local field observables in arbitrary two space-like
separated regions are entangled.
However this method as well as that used in Ref. \cite{unruh}
assumes exact
analyticity and cannot be applied in the presence of a cutoff.

In the present work we consider a gedanken-experiment
for probing entanglement which is not sensitive to a short scale
cutoff. It involves a pair of probes, point-like two-level
systems,
which couple for a finite duration with the field.
The process takes place in two causally disconnected
regions. Since the probes are taken to be initially
non-entangled, and since
entanglement cannot be produced locally,
we will regard the presence of  entanglement in the final state of
the probes as a (lower bound)
measure for vacuum entanglement.

\section{\bf Entropy, Correlations and Entanglement}

We begin with a short review of the relation between entropy,
correlations  and entanglement. Consider a  division of a system
into two sets of commuting degrees of freedom whose combined
Hilbert space can be described by the direct product of Hilbert
spaces ${\cal H}_1\times{\cal H}_2$. The operators ${\cal O}_1$
and ${\cal O}_2$ that act on ${\cal H}_1$ and ${\cal H}_2$,
respectively and thus commute: $[{\cal O}_1,{\cal O}_2]=0$. In a
relativistic theory, the division of space to two space-like
separated regions, implies by causality commutativity of local
observables, and the above Hilbert space structure follows.

We can then distinguish between the two different cases:
a) The system is in a pure state.
and b) the system is in a mixture of
pure states, described by a density operator.
The latter situation can arise, for instance, if our system
constitutes
a sub-system of a larger system whose state is pure. This will
indeed
be the case in our model, as well as that for two separated
regions
in vacuum which do not cover the full space.

Consider first pure states, and for simplicity take
a pair of two level (spin-half) systems.
The Hilbert space contains pure states like
\beq
|\psi\ra = a|\up\ra_1|\up\ra_2 + b |\down\ra_1|\down\ra_2
\eeq
which we call entangled whenever both $a$ and $b$ are non-zero.
More generally, a state is entangled if no local unitary
transformation can
convert the state into a single direct product like $|\ \ra_1|\
\ra_2$.
We observe that entanglement exists if and only if there are
correlations
between local observables: $\la \sigma_1\sigma_2\ra \neq \la
\sigma_1\ra
\la \sigma_2\ra$.

But are these correlations classical or quantum? This question was
answered by Bell and subsequent work. It was shown that
for any entangled pure state,  one can construct an inequality
involving two operator correlations, which are satisfied in a
local classical model but are violated by quantum mechanics.

The existence of correlations means that some information is
stored
in the combined state and cannot be traces by inspecting one
half of the system. A subsystem then
behaves as a mixture. The von-Neumann (or Shannon) entropy
\beq
S_1= -Tr_1 \rho \ln \rho
\eeq
where $\rho = Tr_2 |\psi\ra \la \psi|$, indicates this
lack of knowledge when probing a sub-system.
For pure states, $S_1=S_2$ is non-vanishing if and only if
the state is entangled.
It therefore comes to us as no surprise that entropy can be
viewed as a
quantitative measure of entanglement. The surprise is perhaps
that
for an ensemble of identical states, it is in fact a unique
measure of
entanglement \cite{concentration,sandu+daniel}.

We see that for the case of pure states,  entropy, correlations,
entanglement
(and in a qualitative sense Bell inequalities), are equivalent
descriptions of the same physical phenomena.

The situation differs dramatically in the mixed case.
First how do we define entanglement of a mixed state?
We will define entanglement by saying when a density operator is
$not$
entangled.
A density operator, $\rho_{12}$, is not entangled if
we can find a basis which entails a separable form
\cite{wernerdensity}
\beq
\rho_{12} = \sum_i p_i \rho_{1i}\rho_{2i}
\eeq
with $\rho_1$ and $\rho_2$ as local density operators, and
$p_i>0\ \ \sum
p_i=1$.
We note that $\rho_{12}$ does exhibit non-trivial correlations
since we have
$\la {\cal O}_1{\cal O}_2 \ra = \sum p_i \la {\cal O}_1\ra_i\la
{\cal O}_2
\ra_i$,
but nevertheless the density operator is not entangled, and
describes classical correlations.
Likewise, we note that the entropy function, $S=-\sum p_i \ln
p_i$,
does not vanish for a non-entangled mixed density operator.

To exemplify this, consider the following class of density
operators
\cite{wernerdensity}
\beq
\rho = {1-x\over 4} I + x |{\psi}\ra\la {\psi}|
\eeq
where  $|{\psi}\ra=1/\sqrt2(|\up\ra|\down\ra-|\down\ra|\up\ra)$
is the
EPR-Bohm state.
For any $x\neq0$ the above density exhibits correlations.
However it can be shown that only for $x>1/3$, it is entangled.
It is also interesting to note that, the simple relation between
non-locality described by Bell inequalities and entanglement
breaks down.
In the above case for $x<1/\sqrt2$ there is no violation, though
the state
is entangled.

We see that for the mixed case
entropy and correlations are no longer equivalent to entanglement.
Thus, correlations do not necessarily imply entanglement,
and the von-Neumann entropy is not an appropriate measure of
entanglement.
We will further discuss a possible relation of this issue
with the entanglement interpretation \cite{sorkin} of the
Bekenstein
black-hole entropy \cite{bekenstein} in the last section.

\section{\bf Probing Vacuum Entanglement }

To model the field we shall consider a massless relativistic
scalar field $\phi(\vec x,t)$ in 3 spatial dimensions, driven by
the usual massless free Hamiltonian. We assume that initially the
field is in its ground state. The probe systems will be modelled
by a pair of localized (infinitely massive) two-level systems with
energy gap $\Omega$. The interaction between the systems and the
field will be time dependant but otherwise linear \beqa
H_{int} &=& H_A + H_B \nonumber \\
&=&\epsilon_A(\tau)(e^{-i\Omega \tau}\sigma_A^+ +e^{+i\Omega \tau}
\sigma_A^-)\phi(x_A(\tau),t)
\nonumber \\
&+& \epsilon_B(\tau')(e^{-i\Omega \tau'}\sigma_B^+ + e^{+i\Omega
\tau'}
\sigma^-_B)\phi(x_B(\tau'),t)
\eeqa
where $\tau$ and $\tau'$ denotes the proper time of the probes.
We shall take the coupling functions $\epsilon_A(\tau)$ and
$\epsilon_B(\tau)$,
as non-zero during a finite time $T$ to keep the two probes
causally
disconnected.

entanglement.
for a time

The simplest set up which takes care of this causality restriction
involves a pair of uniformally accelerated detectors that follow
the trajectories
\beq
x_A = -L/2\cosh (2\tau/L) \ \ \
t_A = L/2\sinh(2\tau/l)
\eeq
\beq
x_B= L/2 \cosh(2\tau'/L) \ \ \
t_B= L/2 \sinh(2\tau'/L)
\eeq
Obviously the probes are confined to two causally
disconnected regions; A is confined to the space-time region
 $-x< |t|$ and $B$ to $x<|t|$.

Since the interaction takes place in two
causally disconnected disconnected regions, the field
operators in $H_A$ and $H_B$ commute, and $[H_A, H_{B}]=0$
Therefore, the evolution operator for the system factorizes,
and may be expressed in the interaction
picture as a direct product
\beqa
U  &=&  e^{-i\int\int (H_A(\tau)+H_B(\tau')) d\tau d\tau'}
\nonumber \\
&=& e^{-i\int H_A(\tau) d\tau } \times e^{-i\int H_B(\tau')
d\tau'}
\eeqa
We note that this ensures that $U$
does not generate entanglement between degrees of freedom at
the two causally disconnected regions $x>|t|$ and $x<-|t|$.

Suppose that the probes are initially in their ground states.
 Hence the initial state with the field and probes is
$|\Psi_i\ra =|\down_A\ra|\down_B\ra |0\ra$.
Expanding $U$ to second order in the coupling functions
$\epsilon_i$ ($i=A,B$)
and using the notation
\beq
\Phi_i^\pm = \int d\tau\epsilon_i(\tau)e^{\pm i\Omega
\tau}\phi(x_i(\tau),t)
\eeq
we obtain
\beqa
|\Psi_f\ra
&=&\Bigl[({\bf 1} - \Phi^-_A\Phi_A^+ +
\Phi^-_B\Phi_B^+)|\down\down\ra
- \Phi^+_A\  \Phi^+_B|\up\up\ra
\nonumber \\
&-&i\Phi^+_A {\bf 1}_B|\up\down\ra -
i{\bf 1}_A \Phi^+_B|\down\up\ra \Bigr]|0\ra + O(\epsilon_i^3)
\eeqa
The first term above describes processes where the
initial  state of the probes is
unchanged.
The  second term describes two types of processes,
either an emission
of two quanta, or an exchange of a single quanta between the
the probes.
The final state of the field in this case
is $|X_{AB}\ra\equiv \Phi^+_A\Phi^+_B|0\ra$.
Finally, the last two terms describe an emission of one quantum
either by probe  $A$ or $B$.
In this case the final state of
the field is $|E_A\ra\equiv \Phi^+_A|0\ra$, or $|E_B\ra\equiv
\Phi_B^+|0\ra$, respectively.


Tracing over the field degree's of freedom we obtain to the
lowest order
\beq
\rho=
\left(
\begin{array}{cccc}
1 -C   & - \la X_{AB}|0\ra & 0  & 0 \\
-\la 0|X_{AB}\ra  & |X_{AB}|^2  & 0  & 0\\
0 & 0 &      |E_{A}|^2  &  \la E_B|E_A\ra    \\
0 & 0 & \la E_A|E_B\ra  &  |E_B|^2
\end{array} \right)
\label{density}
\eeq
where $C= 2{\rm Re}\la 0| {\rm T}(\Phi^-_A\Phi_A^+ +
\Phi^-_B\Phi_B^+)|0\ra$,
$|X_{AB}|^2 = \la X_{AB}|X_{AB}\ra$, and we used the basis
$\{|i\ra,|j\ra \}= \{ \down\down, \ \up\up, \ \down\up, \
\up\down \}$.

We noted two types of off-diagonal matrices elements.
The projection of the exchange amplitude on the vacuum, $\la 0|
X_AB\ra$ acts as to maintain coherence between the
$|\down_A\down_B\ra$ and the $|\up_A\up_B\ra$ atom states.
The product $|\la E_A|E_B\ra$ acts to maintain coherence
between the $|\down_A\up_B\ra$ and $|\up_A\down_B\ra$ states.
It is the magnitude of these off-diagonal terms compared
to the magnitude of the diagonal (decoherence) terms which
determine if the density operator is entangled.

For our case of a $2\times 2$ system,
it is known that necessary
\cite{peres} and sufficient \cite{horodecki}
condition for the density operator to be non-separable (and
therefore
entangled)
is that the partial transpose of $\rho$ is negative.
Denoting $\rho_{ij,kl}$ as the matrix
elements  with respect to the basis $|i\ra|j\ra$,
the partial partial transposition takes
$\rho_{ij,kl}\to \rho{il,kj}$.
In our case we obtain the
following two conditions for non-separability.

\beq
|\la 0|X_{AB}\ra|^2 > |E_A|^2 |E_B|^2
\label{first}
\eeq
and
\beq
|\la E_B|E_A\ra|^2 > |X_{AB}|^2
\label{second}
\eeq
When either of these conditions is satisfied $\rho$
is entangled.

The first inequality, (\ref{first}),
 amounts to the requirement that the exchange
process, (which leaves the field in a vacuum state) is more
probable than single quanta emissions
which reduces the coherence of the atoms.
In this case the main contribution comes from states like
$|\down_A\down_B\ra + \alpha |\up_A\up_B\ra$.
Considering the second inequality (\ref{second}),
we note that $\la E_A|E_B\ra$ measures the distinguishablity
of the quanta emitted by either atom $A$ or $B$. Hence
the second inequality demands that
this inner product is larger than the probability $\approx
|X_{AB}|^2$ of emitting two quanta. When the second condition
is met, the main contribution comes from states like
 $|\down_A\up_B\ra + \beta |\up_A\down_B\ra$.

Let us evaluate explicitly the emission and exchange terms
in the first inequality.
The emission term reads
\beq
|E_A|^2 = \int d\tau_A \int d\tau_A'
e^{-i\Omega(\tau_A'-\tau_A)} D^+(A',A)
\label{EA}
\eeq
and the  exchange term
\beq
\la 0|X_{AB}\ra = \int d\tau_A \int d\tau_B
e^{i\Omega(\tau_A+\tau_B)} D^+(A,B)
\label{XAB}
\eeq
where $D^+(x',x)=\la0 |\phi(x',t')\phi(x,t))|0\ra= -{1\over4\pi^2}
( (t'-t-i\epsilon)^2-({\vec x'} -{\vec  x})^2)$
is the Wightman function.
Substituting $x(\tau)$ and $t(\tau)$ one gets
\cite{birrell&davies}
\beq
D^+ (A',A) =- {1\over  4 \pi^2 L^2 \sinh^2[
( \tau_A'-\tau_A -i\epsilon)/L]}
\eeq
and when the points are on different trajectories
\beq    D^+(A,B)
 ={1\over 4 \pi^2 L^2\cosh^2[(\tau_B+\tau_A-i\epsilon)/L]}
\eeq
The integral (\ref{EA}), for the emission probability, can be
performed
by complexifying $\tau_A'-\tau_A$ to a plane and closing
the contour in the lower complex plane. This picks up the poles
$\tau_A-\tau_A' = i\epsilon + i\pi nL$ with $n= -1,-2...$
On the other hand, the contour for the exchange integral
(\ref{XAB})
should be closed on the upper half plane. This picks up the
contributions at
$(\tau_A+\tau_B)=i\epsilon + i\pi (n+{1\over2})L $  with
$n=0,1,2...$
The ratio between the two terms is then
\beq
{|\la 0|X_{AB}\ra| \over |E_A|^2}
= { e^{-\pi
\Omega L /2}{\sum_{n=0}^\infty e^{-\pi n\Omega L} } \over
 \sum_{n=1}^\infty e^{-\pi n\Omega L} } = e^{\pi\Omega L/2}
\label{increase}
\eeq
Therefore (\ref{first}) is always satisfied.
Unlike the previous stationary case, this ratio can become
arbitrarily large, while $X_{AB}$ and $E_A$ become exponentially
small.
The reason for that is that for the hyperbolic trajectories
we can have $\tau\Omega\to \infty$ while keeping $\Omega L$
finite.
By increasing $L$, the emission probability decreases like
$|E_A|^2\sim e^{-\pi \Omega L}$.
However $\la0|X_{AB}\ra\sim e^{-\pi \Omega L/2}$
decreases slower, hence the ratio (\ref{increase})
increases exponentially. It can be shown that similar conclusions
follow
if we switch the interaction only for a finite duration as long
as $\tau\gg 1/\Omega$ is satisfied.

\section{\bf Inertial Probes}

Does the above result have to do with the special effect of
acceleration? To check this we reconsider the problem for the
case of a pair of two stationary probes which are switched
on and off for a duration $T<L/c$ where $L$ is the separation.

Specializing to the case $|\Psi_i\ra=|\down_A\down_B\ra|0\ra$,
substituting $\phi(x,t)$, and
integrating over time
eq. (\ref{first}) can be re-expressed as
\beq
  \int_0^\infty {d\omega\over L} \sin({\omega L}) \tilde\epsilon
(\omega-\Omega)
\tilde\epsilon(\omega+\Omega) >
\int_0^\infty \omega d\omega |\tilde\epsilon (\Omega+\omega)|^2
\eeq
where $\tilde\epsilon(\omega)$
is the Fourier transform of $\epsilon(t)=\epsilon_i(t)$.

The right hand side in the above inequality is independent of $L$
and tends to
zero as $\Omega T\to \infty$. The left hand side depends on both
$T$ and $L$
and decays like $\sim 1/L^2$ for $L>T$.
However for $\Omega L$ not too big,
$\tilde \epsilon(\omega-\Omega)$ has a sharp peak near
 $\omega=\Omega$, which enhances the exchange amplitude.
This suggests that there may exist a finite window of
frequencies around some $\Omega^{-1}\sim T \sim L$, where
(\ref{first})
can be satisfied.

The following plots exhibit
the ratio $X_{AB}/E_A$, for the window function (with $T=1$)
\beq
\epsilon(t) = \left\{   \begin{array}{cc} {\cos^2(\pi t)},
& \ \ \ {\rm for} \ \ \ |t| \le 1/2 \\
              0 , &  \ \ \ {\rm for} \ \ \ |t|>1/2 \end{array}
\right\}
\eeq
as a function of the energy gap $\Omega$ and the separation $L$
between the atoms.

\begin{figure} \epsfxsize=5truein
      \centerline{\epsffile{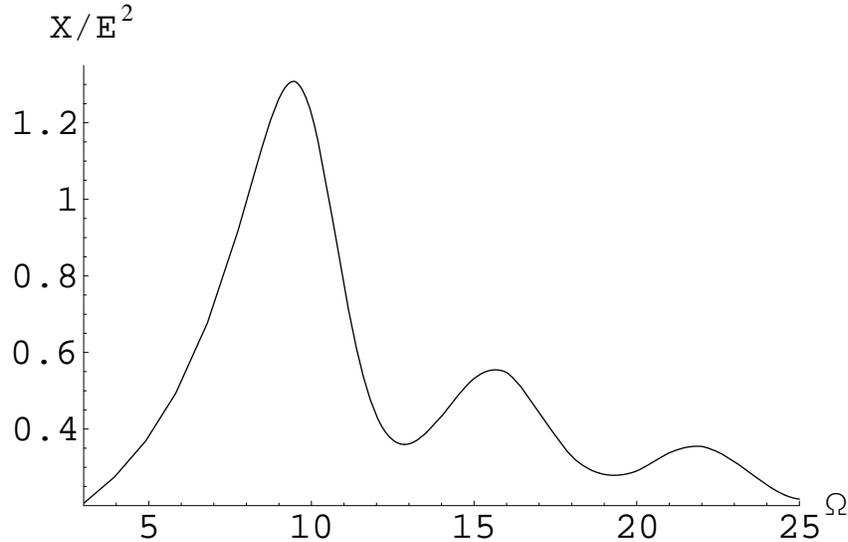}}
\vspace {0.5cm}
  \caption[]{The ratio $X/E^2$,
with $L=T=1$, as a function of the energy gap $\Omega$.  }
    \label{ratio} \end{figure}

\begin{figure} \epsfxsize=5truein
      \centerline{\epsffile{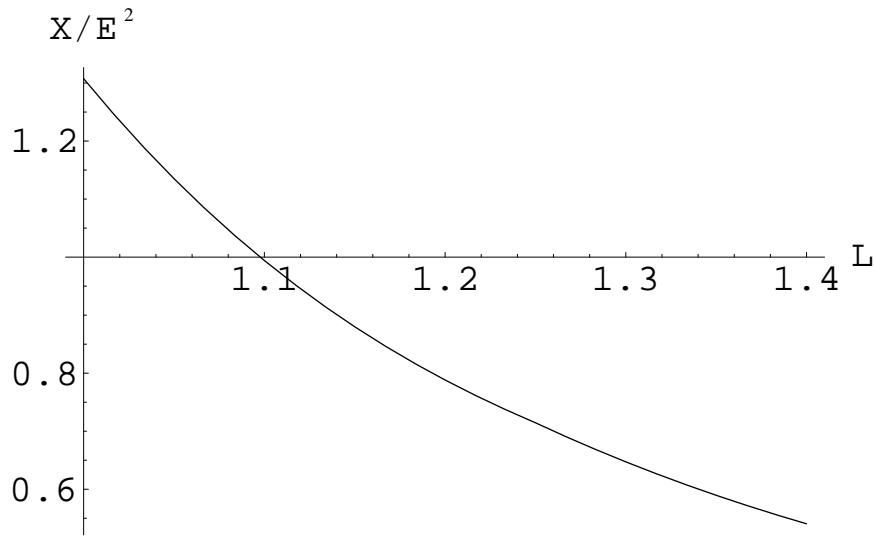}}
\vspace {0.5cm}
  \caption[]{The ratio $X/E^2$, with $T=1$ and $\Omega=9.5$,
as a function of the distance.   }
    \label{ratiol} \end{figure}

It follows from Fig. 1. that Eq. (\ref{first}) is satisfied for
 $8<\Omega <11$.
 For a different distance $L$,
one has to employ atoms with appropriate $\Omega =O(1/L)$.
It follows from Fig. 2 that the spatial region where entanglement
persists,
extends up to $L/T<1.1$. This implies that the maximal
space-like separation between the pair of spacetime regions
that affect the probes  can be extended up to  $L-T \sim 0.1 L$.
This result suggests that the probed must be  "contained" in
the space-like range of a single coherent vacuum fluctuation.
Finally we note that the window functions $\epsilon(\omega)$ act
as
cutoff functions. Hence our result is not sensitive to a cutoff.

\section{\bf Discussion }

We conclude with several comments. We have shown how quantum
correlation, or entanglement, can be extracted to another physical
system. To some extent, such quantum correlations are exploited in
the process of black hole pair creation, but there we have no
access to the quanta emitted into the black hole. Similarly, in
the Unruh effect, the event of thermalization of the detector is
correlated with a flux emitted into the other Rindler wedge
\cite{unruhwald}. In the present process we can control the
regions which we probe, and demonstrate that entanglement persists
even between non-complementary wedges, i.e.,  when the regions are
separated by a finite distance. Nevertheless when the separation
becomes too large the extracted entanglement drops to zero while
the classical type of correlations do not.

We have stressed that entropy is no longer a good
measure of entanglement once the state is not pure,
when for instance the two regions are separated.
The black-hole naturally divides space to interior and exterior
complementary regions, with a combined pure state.
The von-Neumann entropy, coincides in this case with the
entanglement
measure.
How do we renormalize this entanglement entropy?
A naive cutoff seems unjustified, because this will
also truncate the ultra-high modes which are needed in
 Hawking's derivation of black-hole radiation.
On the other hand if we would slightly separate between the
interior and exterior regions (effectively done here by
introducing physical probes) entropy becomes indeed finite, but
would no longer be a measure of entanglement but rather of the
classical correlations.

Finally, we note that
by transforming vacuum entanglement
to pairs of probes or atom-like system, vacuum entanglement
becomes a physical operational quantity.
It has been recently shown that by acting locally on
an ensemble of such generated pairs, one can "purify"
the amount of entanglement and while reducing the number of pairs,
approach gradually to a perfect pure EPR-Bohm pair
\cite{purification}.
The resulting EPR-Bohm pairs
are useful for quantum processes such as teleportation
\cite{teleportation}
or dense coding \cite{dense} which are not possible with the aid
of
classical correlation alone.

\vspace {2cm}

{\bf Acknowledgments\hskip 6pt}

This research was supported by grant 62/01-1 of
the Israel Science Foundation,
by the Israel MOD Research and Technology Unit.

\end{document}